\documentclass[10pt,conference,compsocconf]{IEEEtran}

\usepackage{caption}
\usepackage{comment}
\usepackage[dvipsnames]{xcolor}
\usepackage{cellspace}
\usepackage{array, tabularx, caption, boldline}
\usepackage{graphicx}
\usepackage{cite}

\ifCLASSINFOpdf
\else
\fi
\begin{document}
\pagenumbering{gobble}
%
\title{\textbf{\Large A Survey on Studying the Social Networks of Students}\\[0.2ex]}

\author{\IEEEauthorblockN{~\\[-0.4ex]\large Akrati Saxena \\[0.3ex]\normalsize}
\IEEEauthorblockA{Department of Computer Science, School of Computing\\
National University of Singapore, Singapore\\
Email: {\tt akrati@comp.nus.edu.sg}}\\

\IEEEauthorblockN{~\\[-0.4ex]\large Harita Reddy \\[0.3ex]\normalsize}
\IEEEauthorblockA{Department of Computer Science and Engineering\\
National Institute of Technology Karnataka, India\\
}
\and 
\IEEEauthorblockN{~\\[-0.4ex]\large Pratishtha Saxena \\[0.3ex]\normalsize}
\IEEEauthorblockA{Department of Computer Science and Engineering\\
Center for Advanced Studies, Lucknow, India\\
}\\


\IEEEauthorblockN{~\\[-0.4ex]\large Ralucca Gera \\[0.3ex]\normalsize}
\IEEEauthorblockA{Department of Applied Mathematics,\\ Associate Provost for Graduate Education,\\ Teaching and Learning Commons\\  Naval Postgraduate School, Monterey, CA\\
Email: {\tt rgera@nps.edu}}
\vspace{.1cm}
}


%


\maketitle

\begin{abstract}
Do studies show that physical and online students' social networks support education? Analyzing interactions between students in schools and universities can provide a wealth of information. Studies on students' social networks can help us understand their behavioral dynamics, the correlation between their friendships and academic performance, community and group formation, information diffusion, and so on. Educational goals and holistic development of students with various academic abilities and backgrounds can be achieved by incorporating the findings attained by the studies in terms of knowledge propagation in classroom and spread of delinquent behaviors. Moreover, we use Social Network Analysis (SNA) to identify isolated students, ascertain the group study culture, analyze the spreading of various habits like smoking, drinking, and so on. In this paper, we present a review of the research showing how analysis of students' social networks can help us identify how improved educational methods can be used to make learning more inclusive at both school and university levels and achieve holistic development of students through expansion of their social networks, as well as control the spread of delinquent behaviors. 
\end{abstract}


\begin{IEEEkeywords}
Students' Social Networks; Education; Team Work; Collaborative Learning; Adolescent Behavior
\end{IEEEkeywords}


%
\IEEEpeerreviewmaketitle

\section{INTRODUCTION}


Social networks of students, both physical and online, have been used to understand various phenomena, such as the correlation between the social network position of students and their academic performance, information diffusion in a classroom, collaborative learning and teamwork among students and the emergence of homophily in classrooms~\cite{McPherson2001}. Through the analysis of such students' networks, there can be a progress in the understanding of effectiveness of collaborative learning environments to achieve pedagogical goals~\cite{Nurmela:1999:ECL:1150240.115029}~\cite{MARTINEZ2003353}~\cite{CHO2007309,10.1371/journal.pone.0194777}. Existing research studies the relationship between network dynamics and students' academic and social behaviors. Educators can use this research to ascertain whether enough collaborative learning among students exists, whether there are isolated students who are deprived of help from their peers, and whether there is an effective information flow among the students in a classroom~\cite{Jones2010}~\cite{RAMIREZORTIZ2004175}.

Identifying central students in educational social networks helps in discovering the influential students who spread the information across the network~\cite{5361997}. Using homophily and community detection in such networks, we may also identify the tightly-knit groups of the students that support each other~\cite{RAMIREZORTIZ2004175}~\cite{Smirnov}~\cite{ Brouwer2018}. Dynamic network models explain the evolution of social networks and aid in understanding students' changing behaviors as a social network's links can lead to the development of positive and negative behaviors among the students~\cite{Burk}. These studies help in understanding the spread of various habits such as smoking, drinking and drugs among students~\cite{Steglich2006}~\cite{Lakon2015}~\cite{BurkAlco}.


For the analysis of learning environments, a commonly used method to actively create social network datasets is asking students to nominate their friends, gurus (students from whom they seek help), and adversaries, through filling out questionnaires~\cite{Baldwin}. Also, passive data collection tools like Moodle~\cite{Obadi2010FindingPO} and other collaborative learning systems~\cite{Nurmela:1999:ECL:1150240.115029}~\cite{MARTINEZ2003353} are used, where connections are discovered through students' interactions on online platforms. In both methods, various types of students' social networks can be created, including friendship networks, collaboration networks (where the interaction between the students constitute the collaboration), help-seeking networks (where the students have links with peers whom they ask for academic help), and so on.

With these collection methods in mind, several software programs are currently used for analyzing social networks. UCINET~\cite{borgatti1996ucinet} supports various network analysis techniques including transformation, connectivity measurement, centrality and subgroup identification. The $R$ Package RSiena implements the SIENA method~\cite{siena} and is popularly used to study the relationship between the network and behavior dynamics through stochastic actor based models. The SNAPP tool~\cite{Jones2010} is used to view the real time social network of students based on their online interactions, helping instructors identify the community structures, as well as students who are isolated from others. Gephi~\cite{bastian2009gephi} is another popular open source software that uses 3D rendering for real-time complex networks and supports multi-task modeling. NetworkX~\cite{hagberg2008exploring} is an easy to use Python package that supports a number of graph algorithms.

Having outlined the most commonly used software programs for SNA, we now present a structured review of studies that have used such tools to analyze offline and online social networks of students. The motivations behind the survey are:

\begin{itemize}
    \item Educators are moving away from the traditional teaching methods and are introducing innovative ways to improve learning experience. A good understanding of students' social network structure can help designing effective educational techniques for inclusive learning.
    \item A comprehensive review of the existing methods that use SNA to evaluate students' learning and behavior can provide a thorough understanding of the state-of-the-art in the field of networks and education.
\end{itemize}

Our key contributions through this survey include the systematic partitioning of the work in the area of networks and education into six key fields, highlighting certain limitations in the existing work and suggesting areas for future research.

Section II reviews the existing work under six subsections discussing the use of SNA to study the $(A)$ effectiveness of collaborative learning methods employed by educators, $(B)$ teamwork among student groups, $(C)$ relationship between individual student performance and network structure, $(D)$ extent of knowledge dissemination among students and development of networks for peer support, $(E)$ existence of homophily and communities among students and $(F)$ relationship between network structure, and student habits like smoking and drinking. Section III concludes the survey along with outlining the limitations of existing work and various future directions.



\section{ANALYSIS OF STUDENT SOCIAL NETWORKS}

Social Network Analysis (SNA) can be a good tool in studying the relationships between students forming a collaborative network, between groups of students, as well as individuals themselves as part of a larger network.
Analyzing the network structure indicates not only the extent of learner's benefit from the interactions among the students, but also shows the importance of a particular actor or group in supporting learning in the learning environments modeled by networks. We now discuss the six main aspects related to education that the analysis of students' networks highlights.


\subsection{The Effectiveness of Collaborative Learning Environments}

Early research applying SNA to the field of education studied the interactions in a learning environment furnished by Computer-Supported Collaborative Learning (CSCL). These CSCL environments are learning environments that leverage technology to facilitate learning through computer supported interactions between students such as  online discussion forums and collaborative writing. Analyzing these students' networks helps in evaluating the students' participation in learning achieved through CSCL. For example, university CSCL log files have been used to determine whether each student plays an effective role in the process of learning~\cite{Nurmela:1999:ECL:1150240.115029}. Researchers also created a network structure in which  edges capture the number of sent and received messages between pairs of students on the learning platform. Through the study of betweenness centrality and Stephenson-Zelen information measure~\cite{borgatti1996ucinet}, it can be determined if the communication in the CSCL environment is centralized only around the instructor~\cite{Nurmela:1999:ECL:1150240.115029}. By evaluating whether the interactions in a CSCL environment are instructor-centric, one can perform timely intervention to achieve the goal of more student-student interactions. This evaluation is achieved by measuring ($1$) the level of activity based on nodes' degree, ($2$) the position of students in information exchange network through betweenness and closeness centrality measures, and ($3$) identification of roles of university students~\cite{10.1371/journal.pone.0194777}~\cite{marcos2015despro}. Having fewer student-student interactions implies a poor collaborative environment, and an intervention helps in the emergence of certain students with high betweenness centrality who relay information between the students across the network~\cite{10.1371/journal.pone.0194777}.


Students are a part of different types of networks based on different collaborative activities such as discussion, doubt solving and information sharing. The usage of Basic Support for Co-operative Work (BSCW) software~\cite{bscw} for CSCL shows that doubt solving is the most decentralized activity, whereas the network based on sharing of information has few actors with very high centrality thus implying that information sharing depends a lot on few highly active participants~\cite{MARTINEZ2003353}. It is also of interest to researchers to understand if the formation of collaborative networks in a CSCL community depends on the already existing friendship ties. Formation of such ties can be studied using change propensity and degree centrality~\cite{CHO2007309}. The change propensity measures the extent to which an individual adds more links in her ego network. Students with high centrality in the network of pre-existing ties have a low change in propensity in the CSCL network, whereas peripheral students in the CSCL network tend to show a greater willingness to form connections over time~\cite{CHO2007309}.

An important aspect to understand is whether students start playing a major role in sharing knowledge with the progress of a CSCL based course. The More Knowledgeable Other (MKO) in a CSCL learning environment is the actor with highest knowledge that everyone engages with for guidance~\cite{Vygotskii}. Initially, only the instructor starts as an MKO~\cite{Sundararajan}. Later, the students start becoming more influential in sharing information with their peers. As expected, such an emergence of peer MKOs happens due to increase in the knowledge of the students as the course proceeds and knowledge spreads.

Online Threaded Discussions (OTDs) are often used to enhance peer interactions in university courses. Threads enable students to post their comments asynchronously on a particular topic of discussion and respond to other students' comments. Such discussions can be modeled through directed networks. As expected, high achieving students tend to be the bridges across these discussion groups~\cite{Shaw}. Proper usage of OTDs leads to an increase in the number of students joining the interactions, increase in the number of connections, and a decrease in the average of the closeness centrality of all the students~\cite{Hidalgo}. Decrease in the average closeness centrality indicates lesser delay in information sharing in the network. Meerkat-ED helps to dynamically view students' interactions and identify the central students in each topic's discussion~\cite{meerkat}~\cite{Takaffoli}.

Apart from the above discussed collaborative learning methods, researchers have studied the effectiveness of several other online collaborative learning environments, such as Cooperative Open Learning (COOL)~\cite{Helm}, Peer Assisted Learning (PAL)~\cite{AbdelSalam}, online blogging groups~\cite{Jimoyiannis}, and so on. The use of collaborative learning methods like OTDs successfully increases student interactions. However, it is possible that the students show high participation only when such learning methodologies are introduced for the first time, and may lose interest with repeated usage. Also, the success of such collaborative environments not only depends on characteristics of the social network but also on the motivation and academic abilities of individual students. This is shown in one particular higher education scenario in which even after the adoption of PALs, an individual's academic performance was observed to be comparatively more dependent upon the individual's previous performance than the characteristics of the social network~\cite{AbdelSalam}, highlighting that the usage of PALs may not have been effective in benefiting students through social interactions.

We have discussed how SNA is used to check if collaborative learning methods are able to achieve significant participation of students in the process of learning. More participation implies students ask questions, get their doubts clarified, explain their views and understand other students' views about the topic~\cite{soller1999makes}. We now discuss how the study of network structure helps us evaluate teamwork among students.



\subsection{Studies on the Teamwork Among Students} 

Teamwork is essential among students for performing various activities like group projects, group discussions, or just for sport. Moreover, teams help building long-term connections to possibly support life-long learning. Friendship and communication networks of university students are mostly formed within their teams and high levels of communication within the teams are positively correlated with the team effectiveness, measured by positive outcome~\cite{Baldwin}. Apart from high levels of communication, good team results also depend on balanced communication within a team, measured on the basis of contribution index of each team member, where the contribution index is an indicator of balance in the number of sent and received emails for each individual student within the team~\cite{gloor2008finding}~\cite{gloor2003visualization}. As expected, adversarial relationships within teams are the cause for lesser team satisfaction. One unexpected observation indicates that more workload sharing in a team results in lesser grades. This observation may be explained by the intuition that many successful teams often have a few bright and hard working students who take up most of the workload~\cite{Baldwin}.  


Measures such as cohesion (ratio of the number of mutually positive relationships to the total number of possible relationships), group conflict (ratio of mutually negative relationships to the total number of possible relationships), and degree centrality have been used to understand students' interactions in networks. In particular, based on four types of relationships -- advice, leadership, social and obligation -- research shows that cohesion and centrality are the most important predicting factors for team performance~\cite{YANG2004335}. The edges in the above constructed networks are weighted, with $+1$ indicating a positive relationship and $-1$ indicating a negative one -- it is challenging to capture negative relationships as students may not be willing to reveal such information. To resolve this, the students were presented a questionnaire with only positively stated questions, in which they were just asked to rank their teammates. The lowest ranked student was considered to be having a negative relationship.


The teaching behavior and course structure also impact the evolution of groups and social network formation among students. Based on a study on two sections of undergraduate engineering classes, a less structured class leads to connected groups and some students who are disconnected from others, whereas a more structured class creates a more even distribution of interactions~\cite{Hernandez}. Disconnected students do not interact with their classmates even when teamwork is required. Based on SNA's significance in identifying connected and disconnected groups of students, the instructors can use these results to identify the disconnected students, in order to encourage them to participate in teamwork and to improve the inclusiveness in the projects.

Working in teams helps students in accomplishing course tasks, getting doubts clarified and involving in discussions, thus achieving deeper level of understanding that individual students cannot achieve alone just by attending the course~\cite{sinha2014together}. Teams should have students who play the role of a broker to transfer knowledge between various groups, thus giving the teams access to new and different ideas, information and opinions. An SNA based methodology for forming teams dynamically on Massive Open Online Courses (MOOCs) for achieving tasks assigned in the courses helps in enhancing students' participation, thus reducing attrition from MOOCs through active engagement of students~\cite{sinha2014together}. 

Thus, the level and balance of communication within team, team cohesion, presence of brokers, and absence of disconnected students helps in determining teams' success, both in terms of team grades and the student engagement achieved within the team. In the next section, we change the granularity level to discuss the relationship between an individuals' academic performance and their position in the social network.

\subsection{Academic Performance versus Social Networks}

Analyzing the correlation between students' academic performance and their position in the social network provides interesting insights into the social aspects that make high achieving students different from others. Researchers have focused on studying the relationship of a student's centrality in the network with her academic performance. The centrality in communication networks is a positive indicator of the academic performance of engineering and management students compared to friendship or behavioral networks, and centrality in adversarial networks is negatively correlated with student satisfaction~\cite{Baldwin}~\cite{Mastoory2016}. Advice networks created using data obtained from graduate students in a management course suggest that centrality in advice networks is also a strong indicator of students' academic performance ~\cite{Yang2003EFFECTSOS}. More fine-grained dimensions like the exchange of learning materials, informal communication, and formal study teamwork show that students' centrality in the study environment in university affects both their learning and employability~\cite{divjak}.

A factor that has to be considered while analyzing whether a particular network metric predicts students' academic status is the reciprocal effect, i.e., it is possible that a student with good academics might participate more actively in the network, thus obtaining high centrality values. Hence, it is important to understand the changes in time-varying networks as analyzing a static network may lead to misleading conclusions. The impact of a students' performance on the social network dynamics has been studied through both cross-sectional and temporal networks~\cite{Vaquero}~\cite{Flashman}. High-performing college students tend to establish persistent ties from the beginning of the course itself with other highly performing students. This often leads to the formation of a 'rich' or 'high academic performance' club due to their willingness to collaborate and learn. Thus, academic achievements are a good predictor of the social ties among students having similar academic performance. Also, the students taking advanced coursework often have large ego networks and more interactions in the network~\cite{doi:10.1177/0016986214559639}.

These days, Massive Open Online Courses (MOOCs) are becoming popular and useful resources for education. Such platforms also enable students enrolled in a course to interact via collaborative discussions. Correlation and regression analysis to study the relationships between various social characteristics and the students' online course performance indicate that Eigenvector centrality is one of the most useful predictors for academic achievement, with high-performing students taking up central positions in the network~\cite{Liu_2018}. The studied social characteristics include degree centrality, closeness centrality, betweenness centrality, Eigenvector centrality, PageRank, clustering and hubs.

Spectral clustering on network created on the basis of student participation on university Moodle suggests that similarity in students' behavior is positively correlated with similarity in grades but surprisingly, fails to support the hypothesis that higher centrality causes better performance~\cite{Obadi2010FindingPO}. Analysis of triads and transitivity in student social networks recorded at different times in an introductory biology course, i.e., the network created during their first exam and the network during the second exam showed that the overall number of complete triads indicated the build-up of more study groups in the classroom and increase in the density of the network~\cite{Grunspan2014UnderstandingCT}. Apart from this, there is a significant correlation between performance in the second exam and the students' betweenness and degree centralities. Existing methods can be used to estimate the centrality rank of a student in a class without computing the centrality value of all students~\cite{saxena2017observe}~\cite{saxena2017global}~\cite{saxena2017fast}.

Students are a part of both offline and online social networks, though these networks are often overlapping. Online communication includes communication through mobile and messenger texting, and emails. There is a significant relation between the closeness centrality of the online and offline networks. The offline closeness centrality is significantly correlated with the students' academic performance; however, the correlation is not much in the case of online closeness centrality~\cite{Zhang2008StudentsIA}. The study of online social networks comes with its limitations as the structure and properties of online social networking data are different from the offline collected data, and a rigorous analysis of the data is required while applying the metrics proposed by sociologists and anthropologists \cite{howison2011validity}.


Hence, these studies show that high-achieving students tend to have central position in the network and establish persistent ties with their classmates. Next, we discuss the role of social network in knowledge dissemination and providing peer help to students.

\subsection{Knowledge Dissemination and Peer Support}

Knowledge diffusion is the process by which knowledge spreads over a network. Knowledge can be effectively diffused in students' networks when there is $(1)$ good presence of leaders, $(2)$ optimal network density and $(3)$ existence of subgroups with adequate inter-subgroup connections in a classroom~\cite{5361997}. Degree and betweenness centrality in networks based on emotional, counseling and intelligence relationships between students are used to determine the presence of leaders, and cohesion is used to identify the extent of inter-group diffusion in a university cohort~\cite{5361997}. Bridge students, i.e., the students that constitute the connecting nodes between different groups in a classroom should take up more active roles to improve the diffusion between different groups. As expected, the bridge students are found to be academically outstanding students~\cite{5361997}.

Giving collaboration-demanding assignments to students can lead to the development of an information network that is useful for them throughout their studies. It results in increasing network density, decreasing average geodesic distance, and increasing average degree centrality of students with the progress of a university course~\cite{Han}. The use of student activities requiring intense collaboration also improves information diffusion efficiency and network inclusiveness. The risk of collapse of the collaborative network is also alleviated due to the absence of cutpoint in the evolved network~\cite{Han}. 

Seeking help from peers is often a good way to learn and clarify doubts, especially in advanced courses at the university level. Students who are central in the network constructed on the basis of $getting$ $information$ from peers benefit and learn the most due to gaining of knowledge from others~\cite{hommes2012visualising}. An important question is who the students consult for any help among their peers- their close friends or high-performing students. Though the formation of project groups by university students is highly dependant on the students' pre-existing friendships, among the members of a particular group most students connect to peers who have good knowledge about that subject for help~\cite{Nurminen:2017:FGS:3141880.3141905}. Implementation of an online discussion forum makes it easier for graduate students to seek help by strengthening their social networks. Apart from fueling students' participation in solving problems, such a forum also leads to the expansion of the ego networks of students~\cite{Chao2018StrengtheningSN}.


The relationships between students change due to some special events, but they are stabilized over time \cite{imamverdiyev2010longitudinal}. Similar behavior is observed on Twitter microblogging scenario where the rate of development of connections stabilizes as students start exhibiting more selectivity in forming ties with other students~\cite{5571143}. Application of Stochastic Actor-Based Models (SABMs) for probabilistic analysis shows that high-achieving students obtain more incoming connections with the passage of time implying that high-scoring students gain more attention from their peers, possibly due to the need to seek help in studies.

In this section, we have discussed how the network structure plays an important role in determining the dissemination of knowledge and making it easier to seek help, and how such ideal networks can evolve through collaborative activities. In the next section, we will discuss how SNA is used to study the subcultures in a classroom and the existence of homophily.

\subsection{Study on Subcultures and Homophily Among Students}

The tendency of humans to form relationships with people having similar traits is known as homophily~\cite{McPherson2001}. There is an existence of homophilic connections among high school students, with denser connections among students having similar attributes, such as academic performance, gender, and if they also work while studying~\cite{RAMIREZORTIZ2004175}. Several other longitudinal network data analysis on academic and friendship networks have shown that students mostly form connections with the similarly achieving and same-gender students both for friendship and academic consultation, thus indicating a strong academic and gender-based homophily~\cite{Brouwer2018}.

Do students adapt to the academic performance of their friend circle or try to seek out similarly achieving friends? The observed similarity in academics of friends can be a result of either selection or due to adaptation. Students tend to reorganize their ego networks and form friendships with students having similar academic performance rather than changing their academic performance based on their current set of friends, thus indicating the development of strong homophily. This observation is made through the use of Pearson correlation measure between students' GPA and the average of their direct friends' GPA~\cite{Smirnov}. Girls exhibit more selection phenomenon, which can be explained by the general observation that girls tend to form groups for study far more than boys, thus making academic achievement a major predictor in their friendships~\cite{KRETSCHMER2018251}.


Students can be similar in terms of more than one attribute. However, most studies evaluate different dimensions (of similarities like gender, race, etc.) in a disjoint fashion. For a complete understanding of the impact of homophily on the evolution of the network or other way around, the studies should be performed using multidimensional attributes.  A multidimensional homophily study on adolescent networks shows that the connections between individuals having more than one similar attribute might not evolve further~\cite{Block2014}. This is explained with the idea that students may usually try to make friends with people who have different thoughts and knowledge, thus seeking variety in their friends' circle.  

How does SNA help in identifying the sense of community in students? Degree and closeness centrality are positive indicators of the sense of community in a student whereas betweenness centrality shows a negative correlation. This conclusion is obtained from the evaluation of relationships in evolving students' social networks and the feeling of community (social and learning) in those students using Classroom Community Scale (CCS)~\cite{Dawson}~\cite{Rovai}. However, this study only considers communication through online discussions.

Homophily based groups are helpful for students to feel inclusiveness, but at the same time, they have several drawbacks. The poorly performing students might form groups among them and it may lead to an inverse effect on their academic performance, therefore institutions and teachers should focus on the disruption of such groups and formation of more versatile groups. Strong homophily in a classroom is also an indicator of segregation based on racial or ethnic terms; development of inter-group ties can help develop better inter-racial and ethnic relations among students~\cite{Hajdu}. Research shows that strong homophily in a classroom can be overcome by creating more connections between separated groups based on performance or legislation. For example, one study has shown that minority groups can have better inter-ethnic relations by performing better in academics. That is because by performing well the minority students tend to get more attention and therefore more friends and lesser adversaries from the majority groups~\cite{Hajdu}. SNA has also been used to study the impact of affirmative actions, such as reserving seats for underrepresented community students, motivating more women to study sciences and technologies, organizing events to bridge the gap of two communities, etc~\cite{saxena2015social}. A mathematical analysis of the seat reservation system for backward class students in Indian academic system showed that such affirmative actions reduce the gap of backward and upper-class students with time~\cite{saxena2015social}. This work is based on a very basic mathematical modeling of the network and needs to be extended by considering more real-life parameters. The survey included in the research also showed that the opinion of upper class students changes about backward class students once they meet more of such students~\cite{saxena2015social}.

In this section, we discussed how SNA is helpful in studying the community, groups, and subculture formation among students based on similar attributes like race and gender. Another important factor that influences the evolution of ties is similar habits. In the next section, we will discuss the impact of the behavior of friends on a person and friendship formation based on behavioral attributes.

\subsection{Studies on Adolescent Behavior}

Friendships can influence a student's behavior in several ways, ranging from improving academic performance to getting an antisocial behavior. Students tend to have connections with peers having similar behavior, including aggressive behavior~\cite{Cairns1988}. For the majority of adolescents, increase in individual importance in the social network leads to increased aggression, except for very highly central students who do not need to resort to aggression to improve their social standing, with the social standing measured through betweenness centrality~\cite{faris2011status}. A rise in the average aggression of friends also leads to a rise in the aggression of an adolescent \cite{faris2011status}. There is a significant impact of friends' delinquent behavior on an adolescent and this impact is beyond that of the impact of school involvement. Study of structural characteristics of the network and behavioral dynamics model explains such a correlation between changing behaviors of adolescent students and their changing network connections~\cite{Burk}.

\begin{table*}[tbhp!]
\begin{center}
\caption{\label{tab:t1}COMMONLY USED NETWORK SCIENCE CONCEPTS TO ANALYZE STUDENTS' NETWORKS}
\label{fake}
\begin{tabular}{|l|c|}
\hline
\textbf{Network Concept}&  \textbf{Applications}\\
\hline

Degree Centrality&Find centrally positioned high performing students~\cite{Yang2003EFFECTSOS}~\cite{Grunspan2014UnderstandingCT}, willingness of students to add more ties in CSCL environment~\cite{CHO2007309}\\

&team performance~\cite{YANG2004335}, development of information network~\cite{Han}, sense of community in students~\cite{Dawson}\\

\hline

Closeness Centrality&Find students with good academic performance~\cite{Zhang2008StudentsIA}, delay in information sharing in network~\cite{Hidalgo}, sense of community in students~\cite{Dawson}\\

\hline

Betweenness Centrality&Determine if the communication is centralized around the instructor in a CSCL environment~\cite{Nurmela:1999:ECL:1150240.115029}, students who relay information across\\

&the network~\cite{10.1371/journal.pone.0194777}, students who perform well in exams~\cite{Grunspan2014UnderstandingCT}, the presence of leaders~\cite{5361997}, increase in aggression~\cite{faris2011status}\\

\hline

Stephenson and&Find centrally positioned high performing students in communication networks~\cite{Baldwin}, determine if the communication\\

Zelen Centrality&is centralized around the instructor in a CSCL environment~\cite{Nurmela:1999:ECL:1150240.115029}\\


\hline

Eigenvector Centrality&Find centrally positioned high-performing students~\cite{Liu_2018}\\

\hline

Cohesion&Predict performance of student teams~\cite{YANG2004335}, find the extent of inter-group information diffusion~\cite{5361997}\\

\hline

Network Density&Find if knowledge is effectively diffused in classroom~\cite{5361997}, development of information network due to collaborative assignments~\cite{Grunspan2014UnderstandingCT}~\cite{Han}\\

\hline

Strongly Connected&Study the improved flow of information due to decrease in the number of components through the use of threaded discussions~\cite{Hidalgo}\\

Components&\\

\hline

Triads&Study the formation of study groups in a classroom~\cite{Grunspan2014UnderstandingCT}\\

\hline

Bridges&Find students who can play an important role in improving information diffusion in classroom, and are usually academically outstanding~\cite{5361997}\\

\hline

Geodesic Distance&Study the development of information network and how information diffusion improves~\cite{5361997}\\

\hline

Stochastic Actor&Study the phenomenon of influence and selection of friends in terms of smoking~\cite{Lakon2015} and drinking~\cite{BurkAlco}~\cite{Light}~\cite{Osgood} habits,\\

Based Model& physical activity levels~\cite{Simpkins}, academic achievements~\cite{Flashman} and junk food consumption~\cite{Hayejunk} among students\\


\hline

Homophily&Evaluate the tendency of students with similar attributes forming friendship ties with each other~\cite{RAMIREZORTIZ2004175}~\cite{Brouwer2018}~\cite{Block2014}, \\


&racial and ethnic segregation in classroom~\cite{Hajdu}\\

\hline

\end{tabular}
\end{center}
\end{table*}

Students acquire several habits and behaviors through the influence of their friends. Such relationships between network dynamics and behavioral changes are often studied through SABMs~\cite{SNIJDERS201044}. Students modify their smoking habits to match their friends, thus corroborating that influence from their network increases smoking among youth. This is observed through the study of dynamic networks based on three measures: smoking alter, smoking ego, and smoking similarity among students~\cite{Lakon2015}. Similar methods are used to study the impact of social networks on alcohol consumption by the youth. Adolescents often select friends with similar drinking behavior and start consuming alcohol if a large number of friends already have the habit of consuming alcohol; alcohol onset can be seen as a diffusion process in the students' network~\cite{BurkAlco}~\cite{Light}. Moreover, adolescents often get attracted to drinkers and try to be friends with them due to the teenage culture of giving high status to drinkers as shown through the SABMs analysis~\cite{Osgood}. The selection of friends is also heavily influenced by levels of marijuana use based on the study performed in two schools \cite{Haye}. However, the phenomenon of non-users trying marijuana for the first time because of their friends' influence is observed only in one of the schools. Such analyses can help in identifying schools where youth drug interventions are required to prevent potential marijuana use.

In understanding adolescents' health, students gradually adjust to obtain a Body Mass Index (BMI) and physical activity levels similar to that of their friends, as shown by SABMs in a study of the relationship between longitudinal adolescent friendship networks and their physical activity~\cite{Simpkins}. The converse is also true, i.e., there is a tendency of students to connect with students having similar BMI and physical activity levels, based on homophily as discussed in the previous section. Moreover, apart from a few personal attributes, an adolescent's junk food consumption is also largely influenced by her friendship network~\cite{Hayejunk}. Such SABM based studies can be used to understand student behavior and intervene in case of disorders faced by them. Based on these findings, educational institutions may design appropriate policies and activities to overcome these patterns, making the students aware of this data and motivating a better decision while choosing friends.

Another research problem is the correlation between students' networks and the prevalence of cheating. Empirical evidence suggests that formation of tightly knit groups or $cliques$ among students, and weak ties with the faculty is positively correlated with the adoption of unethical behavior~\cite{brass1998relationships}~\cite{hutton2006understanding}. Friendship networks can be used to design effective seating arrangement strategies for exams to prevent cheating. Optimized solutions for lattice-based student placement layouts, with less lattice links matching with actual friendship network links can be obtained through the use of genetic algorithms~\cite{topirceanu2017breaking}.


The section elaborated on the relationship between dynamics of network ties and adolescent behavior including aggression, smoking, and consumption of alcohol, marijuana and junk food. We now conclude our review and elucidate the possible directions of research in future.

\section{Conclusion and Future Directions}

Analysis of networks of students constructed on the basis of different types of relationships helps us understand students' behavior, information diffusion, and acquisition of various habits like smoking and drinking based on network structure. Research shows the correlation between the academic performance of the students and their position in the network. These observations can be used to design correcting behaviour, alternate teaching methods, group activities, and policies that impact students. A limitation of the studies carried out on students' networks is that generalization drawn from the observations on small groups of students may not always be correct. Another issue is that self-reporting on adversarial relationships, and habits like smoking and drinking may not give accurate information as it is subject to students' bias and hesitation. There is also a need to increase the frequency of evaluation of students' networks using the different frameworks. Evaluation should also be done a significant amount of time after the introduction of new learning methodologies to check if students' participation remains persistent.
Currently, there is little research that studies combined teacher-student social networks. These mixed networks may provide more information than what only students' or teachers' social networks provide individually. Such combined networks can be explored for studying the long-lasting teachers' impact on the students, and their influence in the formation of groups among students. We can also determine students who are not well-connected to the faculty and encourage them to interact more. Also, real-life scenario based case studies need to be conducted to evaluate teacher-student networks in terms of prevalence of unethical practices among students like cheating. 

Student segregation based on racial, ethnic or class homophily is detrimental in nature. Future research can compare the changes in homophily in students' networks over time at different universities that have different amount of diversity and that have different or no programs for encouraging intermingling. This can help in identifying the best methods for disrupting segregation in higher education.




Another network science based learning tool is CHUNK learning methodology that takes into consideration the different learning abilities and skills of students, and uses network-based approach to represent content and suggest learning modules to students~\cite{chunklearning}. Such tools can complement existing collaborative learning techniques to furnish a personalized learning experience and enable all students to participate proactively in learning. 

Analysis methods should be easy to implement in real life scenarios, with lesser time spent by students like in filling questionnaires. Certain studies use rigorous questionnaires and surveys which are time-taking to implement on a regular basis in universities or schools. We suggest that future research should focus on automatic data collection using online platforms like Moodle and discussion forums, which have already been utilized by some studies, along with integration of student ties on online social networking sites. Software should be developed to analyze networks constructed from the data to examine student participation and identify disconnected students so that the faculty can identify appropriate corrective steps. We believe that teachers should be able to use the SNA tools easily to continuously monitor the students' social networks to bring awareness to possible appropriate interventions that improve the interactions and knowledge diffusion in a classroom. We promote social network analysis as a tool to complement easily observable behavior, but not to replace the observations captured through human teacher-student or student-student interactions. 
\section*{Acknowledgments}
Ralucca Gera would like to thank the DoD for partially funding this work.






%
%
%

\bibliographystyle{IEEEtran}
\bibliography{mybib}

\end{document}